\documentclass[aip,preprint]{revtex4-1}

\usepackage{hyperref}
\usepackage{graphicx}
\usepackage{amssymb, amsmath}
\usepackage{epsfig}

\begin{document}
\title {A powerful computational crystallography method to study ice polymorphism}
\author{M. Cogoni}
\email[Corresponding author. E-mail:]{mcogoni@crs4.it}
\author{B. D'Aguanno}
\author{L.N. Kuleshova}
\author{D.W.M. Hofmann}

\affiliation{
CRS4, Center for Advanced Studies, Research and Development in Sardinia,
Edificio 1 - Parco Scientifico e Tecnologico della Sardegna,
09010 PULA (CA - Italy)
}

\begin{abstract}

Classical Molecular Dynamics (MD) simulations are employed as a tool to
investigate structural properties of ice crystals under several temperature and
pressure conditions. All ice crystal phases are analyzed by means of a
computational protocol based on a clustering approach following standard MD
simulations. The MD simulations are performed by using a recently published
classical interaction potential for oxygen and hydrogen in bulk water,
derived from neutron scattering data, able to successfully describe complex
phenomena such as proton hopping and bond formation/breaking.
The present study demonstrates the ability of the interaction potential model to
well describe most ice structures found in the phase diagram of water and to
estimate the relative stability of sixteen known phases through a cluster
analysis of simulated powder diagrams of polymorphs obtained from MD simulations.  
The proposed computational protocol is suited for automated crystal structure
identification.
\end{abstract}

\maketitle

\section{Introduction}

Many attempts have been performed to reproduce the complex phase diagram of water,
in particular the ice polymorphs, by means of classical Molecular Dynamics (MD)
\cite{malenkov_zheligovskaya,guillot2002reappraisal,sanz-vega,bartok_2005}.
Since MD is essentially based on the evolution of an exact configuration of
atoms in phase space, water is usually treated in a peculiar way because of the
non purely classical behavior of hydrogen.
Most classical interaction potentials treat the
water molecule as a rigid body composed of three single point charges (i.e. SPC),
not allowing, or significantly reducing, the degrees of freedom associated to the 
hydrogen atoms. This approximation is acceptable only when the focus is not on
the hydrogen atoms themselves, with water molecules supposed to be perfectly
stable and no proton dynamics expected. Some other potentials are based on more
complex approximations which add bond flexibility to the water molecule (i.e. SPC 
and TIP4P variants).
However, all such classical force fields are unable to quantitatively
describe hydrogen diffusion. A variety of potentials for water is presently
available, and for a list of them see the review by Guillot
\cite{guillot2002reappraisal} and references therein. In the literature, the most 
used potentials for ice simulations are the TIP4P and SPC/E potentials
\cite{sanz-vega, bartok_2005}. 

To date, the most systematic MD investigations on the relative stability of ice
phases are the work of Zheligovskaya \cite{Zheligovskaya}, based on the
potential of Poltev-Malenkov \cite{poltev-malenkov}, and that of Baranyai et al.
\cite{bartok_2005}, which uses SPC/E and TIP4P water models. 
From such investigations, it is clear that the used
model potentials are not able to reproduce all sixteen known crystal structure
of ice. Inaccuracy of current potentials, and the rigidity imposed on the water
molecule are considered the main reasons for this failure \cite{Zheligovskaya,
parrinello, bartok_2005}, confirming that for an adequate description of the ice
and/or water systems by classical MD, the quality of the potential is of prime
importance. 

Earlier, some of the authors of this study  introduced a new reactive force
field for water (RWFF)\cite{hofmann-r-potential}, which was derived from neutron
scattering data. This force field, while of classical nature, allows proton
diffusion by a breaking/reforming mechanism of O-H bonds in water molecules, in
acids and in hydronium ions, and correctly describes the proton dynamics and the
structure of bulk water. The RWFF has been extensively and successfully applied
to the theoretical investigation of conductivity in polyelectrolyte membranes
for fuel cells\cite{hofmann_fc, hofmann_conduct}.

In this study, and by using the RWFF potential, several MD
simulations have been carried out by encompassing the range of pressure, $P$, and
temperature, $T$, needed to analyze all crystal phases of water. The so obtained
crystal structures are then ready to be fully characterized in their structure.

A large amount of publications is also dedicated to the development of efficient
methods  able to characterize and classify the different ice structures.
Extensive studies on ice classification have been made by Chaplin\cite{m_chaplin}
and Malenkov\cite{malenkov2009}. The used approach is based on the detection of
different local topologies of the bond network such as proton disorder/order,
ring sizes, helices, angles, and ring penetration. The crystal structures are
then compared pairwise according to the different properties. However, such a
procedure is not well suited for an automated classification analysis. 

Other complex classification criteria
have been introduced to analyze water structures and ice polymorphs
obtained during MD simulations. Carignano et al.\cite{carignano} looked at the
time average of the number of H-bonds around O-atoms to detect different phases
of water and ice Ih. Moore and Molinaro\cite{moore_2009} developed an algorithm
which focuses on the first coordination sphere of each oxygen, where any deviation
from perfect tetrahedrality classifies water molecules as belonging to the
liquid or to the solid phase in two clusters. This algorithm has been extended
to recognize water molecules belonging to Ih or Ic structures \cite{moore_2010}.

An alternative method is based on the analysis and classification of the crystal
structures by means of their clustering in the reciprocal space
\cite{hofmann-similarity}. Ice crystal structures can hardly be compared in the
direct space, since the presentation of the unit cell is not unique, and small
changes of the lattice can easily be missed. However, a Fourier transform of the
real coordinates gives an unique presentation for each phase, while the
multiplication of the Fourier transform with the scattering factors results in
the powder diagrams. In the present study, the powder diagrams are generated for
samples obtained in MD simulations and then used to classify the ice crystal
structures.
 
Clustering techniques rely on the definition of a similarity index capable of
reducing to a single number many relevant geometrical properties.
Such techinques, have already been defined and employed for the estimation of the
number of local minima within the energy landscape of crystal structures
\cite{hofmann-minima} and for automated crystal structure determination
\cite{hofmann-automated}.
In this way, a quick and univocal detection of manifold and isostructural
crystal structures in extensive data sets \cite{data-mining} was possible.
In the present work, the similarity index introduced in Hofmann et al.
\cite{hofmann-similarity} has been employed and more details on its definition can
be found there. For our purposes, we recall that this index is based on comparing
integrated powder diffraction diagrams instead of the powder diagrams themselves.
The index is defined as the mean difference between two normalized
integrated powder diagrams and is proportional to the area between the two
curves. In contrast to traditionally used similarity indices, this method is
valid for comparisons where large deviations of cell constants are present.

In this work the method is applied to identify the different crystal structures
of ice, outcomes of long MD simulations over a broad range of temperature and
pressure.

The paper is organized as follows: In Section 2 the phase diagram of water
and a classification of all known ice crystal structures are presented.
Section 3 is dedicated to the illustration of the crystal clustering analysis
and of the proposed computational crystallography protocol.
The protocol is established with the aim to accurately verify the ability of the
RWFF potential of reproducing the stability of the ice crystal structures found
in each point of the thermodynamical phase diagram.
Molecular Dynamics details needed to perform the simulations are reported in
Section 4. The results of the computational crystallography protocol are shown
and discussed at length in Section 5. The last Section is dedicated to the 
conclusions. 

\section{Taxonomy of ice structures}
\label{Characterization}
\begin{figure}
\centering 
\includegraphics[width=12cm]{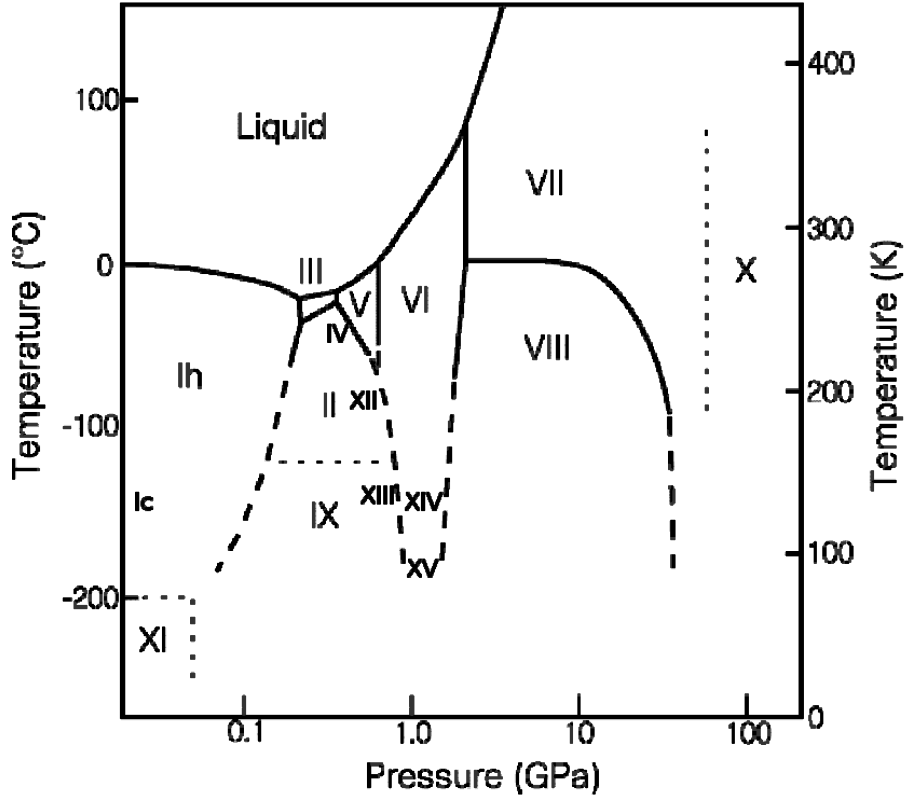}
 \caption{Phase diagram of water \cite{lobban1998structure}. Ice IV, XII, XIII,
XIV, XV (not included in the original diagram) are placed near their
known stability regions.}
\label{phase-diagram}
\end{figure}

At present, sixteen crystal morphologies of ice are known within the water phase
diagram over a very broad range of temperature and pressure. The different ice 
phase regions and structures are shown in Figure \ref{phase-diagram}. 
Most of the ice modifications are characterized by specific regions of stability,
while some others (Ic, IV, XII) occur exclusively as metastable states. 

Some crystallographic information concerning ice polymorphs is summarized in
Table \ref{ices}.  Water crystallizes in different space groups and the density
of each structure varies over a broad range from $0.930$ to $2.785$ g/cm$^3$
(sixth column of Table \ref{ices}).
The crystallographic densities are reported in this table and they can be
slightly different from the experimental values due to the presence of
crystal defects in real samples (e.g. vacancies, interstitials, etc.) The number
of molecules in the asymmetric cell ranges from 1/2 to 7 (column Z'), while the
number of molecules in the unit cell goes from 2 up to 28 (column Z).
According to the H atom positions (water molecules orientation), proton ordered
and proton disordered phases are defined (column o/d). Some of these phases
differ just because of proton order/disorder \cite{malenkov2009}.

Table \ref{ices} reports the configurational energy (sum of RWFF and of the
electrostatic terms) for most of the structures obtained in this work and a
comparison with two widely used force fields for water.

In the structures of all known polymorphs of ice, with the only exception of
ice X, each water molecule forms four hydrogen bonds with neighbor molecules.
In ice X each proton does not belong uniquely to a single oxygen atom, but it
is shared between two oxygens. This fact leads to a quantum behavior that can
only be approximately described by means of a standard MD\cite{parrinello-iceX}.

The ``low pressure'' phases (Ih, Ic, XI) are characterized by a
nearly ideal tetrahedral environment of the oxygen atoms, while ``medium and
high pressure'' phases (II-IX, XII-XV) show a more or less distorted
tetrahedral coordination of the oxygens. 

\begin{table}
\center
\caption{Characteristics of ice structures. $\Delta U_{pot}$ is computed by subtracting
from the energy of each structure the energy of the most stable (ice XI.) Data
relative to TIP4P and SPC/E taken from Vega et al.\cite{vega-gdr1} and Martin-Conde
et al.\cite{vega-gdr2} }
\begin{tabular}  {|l|l|l|l|l|l|l|l|l|l|l|l|l|}			
\hline
structure  &o/d& space group &  Z'     & Z  & $\rho$ (g$\cdot$ cm$^{-3}$) &
$\Delta U_{pot}$(kJ/mol) RWFF & TIP4P & SPC/E\\
\hline
Ic 	&d& $Fd\overline{3}m  $	& 0.5 	&8	& 0.933  & 1.31  &1.92 &2.13 \\
Ih	&d& $P6_3/mmc 	      $	& 1 	& 4  	& 0.926  & 9.65  &7.19 &7.53 \\
II	&o& $R\overline{3}    $	& 2	& 12 	& 1.195  & 6.67  &4.18 &2.13 \\
III	&d& $P4_12_12         $	& 1.5 	& 12 	& 1.160  & 9.76  &8.49 &8.83 \\
IV	&d& $R\overline{3}c   $	& 2   	& 16 	& 1.275  & 8.69  &5.52 &6.02 \\
V   	&d& $ C2/c            $	& 4   	& 28 	& 1.233  & 15.36 &7.78 &8.45 \\
VI  	&d& $ P4_2/nmc        $	& 1.5 	& 10 	& 1.314  & 14.17 &8.20 &9.20 \\
VII 	&d& $ Pn\overline{3}m $	& 0.5 	& 2  	& 1.591  & 27.50 &7.78 &8.45 \\
VIII	&o& $ I4_1/amd        $	& 1   	& 8  	& 1.885  & 21.61 &8.95 &10.12\\
IX  	&o& $ P4_12_12        $	& 1.5 	& 12 	& 1.160  & 2.26  &4.68 &4.31 \\
X   	&nm&$ Pn\overline{3}m $ & 0.5 	& 2 	& 2.785  & 171.70&-    &-    \\
XI  	&o& $ Cmc2_1          $	& 1.5 	& 8  	& 0.930  & 0.00  &0.00 &0.00 \\
XII 	&d& $ I\overline{4}   $	& 1.5 	& 12 	& 1.301  & 10.54 &9.45 &10.21\\
XIII	&o& $ P2_1/c          $	& 7   	& 28 	& 1.247  & 8.16  &3.76 &3.22 \\
XIV 	&o& $ P2_12_12_1      $	& 3   	& 12 	& 1.294  & 6.91  &3.47 &4.18 \\
XV  	&o& $ P\overline{1}   $	& 5   	& 10 	& 1.328  & 7.09  &-    &- \\
\hline
\end{tabular}
\label{ices}
\end{table}

\section{Crystal clustering analysis and computational crystallography protocol}

To verify the accuracy of the RWFF potential in describing all the sixteen ice
structures reported in the previous Section, the following ``computational
crystallography protocol'' is adopted:
\begin{enumerate}
	\item Make a choice of a thermodynamic point;
        \item Perform MD simulations by starting from the ideal crystal structures,
        and store the final thermalized crystal structures;
        \item Compute the powder diffraction diagrams for the initial and for the
	equilibrated ice structures, then evaluate the corresponding similarity
	indices;
        \item Cluster the structures, and draw the dendrogram as obtained from
        the average linkage distance.        
\end{enumerate}
When the final structures of the MD simulations cluster with the starting ideal
crystal structures, the force field is able to mantain structural stability.

In this Section and to give an example of such a protocol, the dendrogram 
obtained for all ice structures from published theoretical crystal data
\cite{ICSD} is evaluated.

\begin{figure}
\centering
\includegraphics[width=0.65\textwidth]{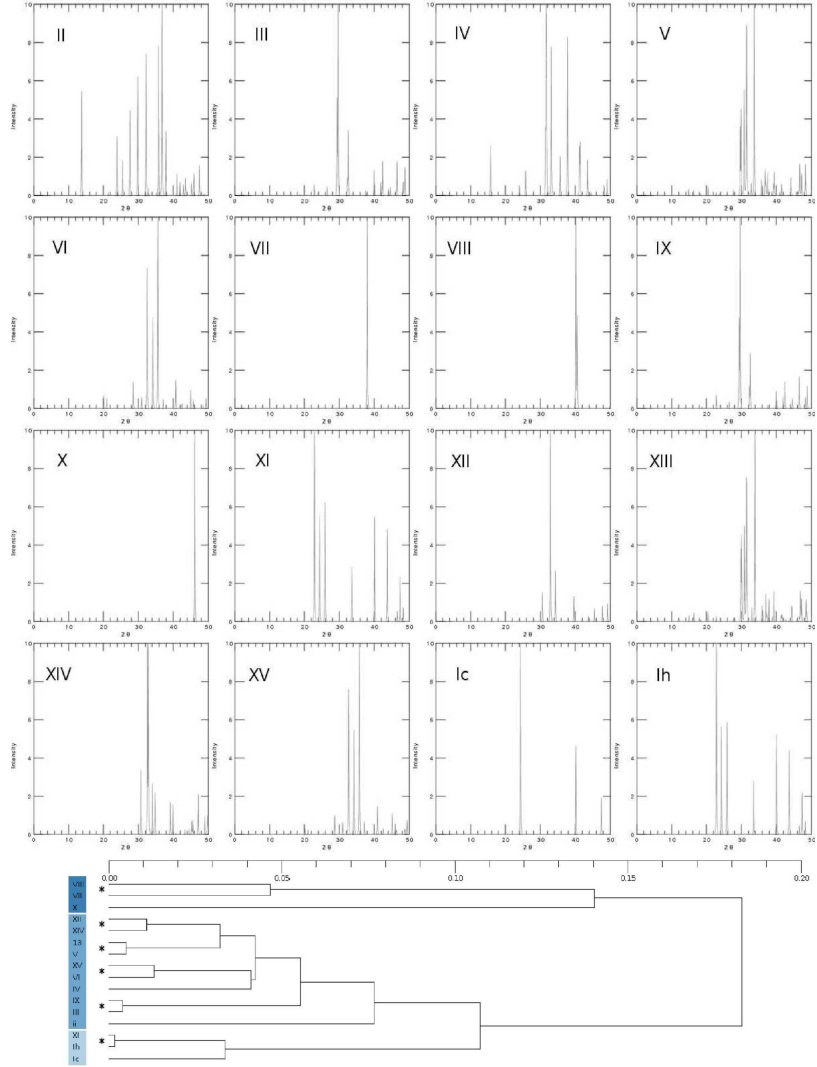}
\caption{Powder diagrams of the ideal crystal structures and their
dendrogram. Asterisks indicate ordered-disordered crystal couples (darker shades
of blue correspond to stability at higher pressures.)} 
\label{dendrogram}
\end{figure}

Since in this case step 1 and 2 are not required, the protocol starts from step 3.
For the step 3, the calculation of powder diffraction diagrams and of their
clustering for all
ice structures was done with {\it FlexCryst} software \cite{flexcryst}.
Here, it is stressed that the clustering approach helps to visualize, in a single
picture, the complex relations existing within the crystal structure set, and to
intuitively understand the influence of temperature and pressure.
The results of step 4 are reported in Fig. \ref{dendrogram}, where the powder
diagrams of the ideal crystal structures (upper part), and the result of the
clustering operation on them (lower part), are shown.

The dendrogram reveals three structural clusters, making visible a
taxonomy ruled by pressure effects: The crystal structures split in a ``low
pressure'' cluster with the polymorphs Ih, Ic, XI; a ``medium pressure'' cluster
with the polymorphs II-VI, IX and XII-XV; and a ``high pressure'' cluster with ice
VII, VIII, X. Within these clusters, and in agreement with the data summarized
in a recent paper by Malenkov\cite{malenkov2009}, one can easily spot the proton
ordered-disordered couples of structures (marked by asterisks in the dendrogram),
even if they belong to different space groups: Ih-XI, III-IX, V-XIII, VI-XV,
VII-VIII, XII-XIV.

In the rest of the paper the protocol will be used in its full extension.

\section {Molecular Dynamics simulations}
\label{protocol}
\begin{figure}
\centering
 \includegraphics[width=0.6\textwidth, angle=270]{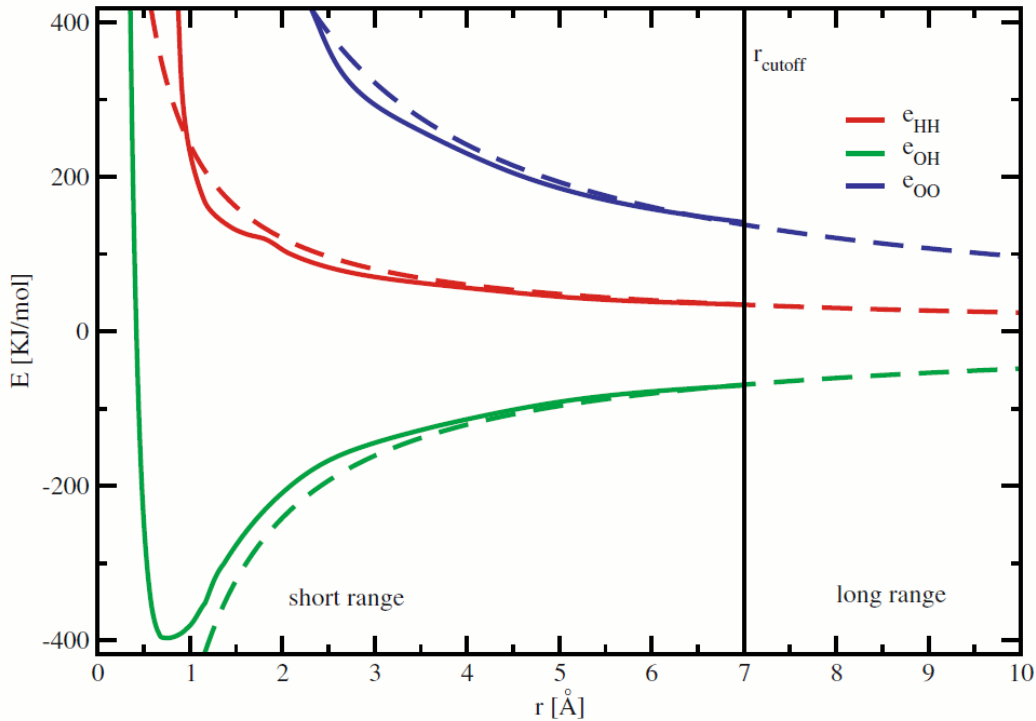}
\caption{The graphical representation of the RWFF potentials. The short range
potentials are shown in full lines. After the cutoff at 7 \AA \,  the potentials
are continued by the Coulomb terms (dashed lines).} 
\label{rwff}
\end{figure}

For the MD simulations, it has been used a customized version of DL$\_$POLY V.2
\cite{DLPOLY} modified in order to include the reactive water force field,  RWFF
\cite{hofmann-r-potential}. The RWFF potential, which was derived to reproduce
correctly liquid water neutron scattering data over a broad range of temperatures
and pressures \cite{hofmann-r-potential}, and which has been employed in the
present study, is shown in Fig. \ref{rwff}. It is only available in a tabular
form which is automatically interpolated by DL$\_$POLY.

\subsection {Initial configurations}
\label{start}
For each ice structure,  a supercell with 1680 water molecules has been generated
by repeating the unit crystal cell along the three spatial dimensions.
Hereby, the supercell was constrained to an approximately cubic shape.
The atomic coordinates were taken from the Inorganic Crystal Structure Database,
(ICSD) \cite{ICSD}. The size of the supercell is a
compromise to minimize finite-size effects and CPU-time. The number 1680 is the
least common multiple of the different numbers of molecules in the unit cells
of the polymorphs. Therefore, such a supercell can be used for all of them.
For the cases in which the hydrogen positions were experimentally undefined, the
hydrogen atoms have been placed by following the Bernal-Fowler rules
\cite{bernal-fowler}: This freedom in configuration building is due to the use
of the RWFF model which allows to place the hydrogen atoms anywhere in the lattice.

\subsection {Details of MD simulations}
The initial configurations, generated as described in the above
subsection, were then equilibrated by MD run
within the isothermal-isobaric ensemble (NPT), in which the number of
atoms, the pressure and the temperature have been kept constant during the
simulation. 
Real space cutoff for the RWFF interaction was set to $7.0$ {\AA} while for the
long range Coulomb part, a full Ewald summation was employed.
An anisotropic Nos\'e-Hoover barostat and thermostat has been employed with
a time constant of $0.4$ ps. In this way the simulation box is free to
optimize both its volume and its shape. Pressure and temperature do not show any
drift or long period oscillations during the simulations.
The timestep was set to $0.1$ fs, which is smaller than common
MD simulations for water. This small timestep is necessary to describe accurately
the fast vibrations of the protons and to conserve total energy over long time scales.
Each simulation was evolved for $1.0\cdot10^{6}$ steps using a Verlet algorithm
\cite{verlet}, so that the total time reached 100 ps.

After the equilibration run in NPT ensemble, every sample was evolved for
further 100 ps ($1.0\cdot10^{6}$ simulation steps) in the microcanonical ensemble
(NVE) with constant number of atoms, volume and internal energy.
During the NVE run the energy has always been conserved within $0.1\%$ over 
$1.0\cdot10^{6}$ steps.
The NVE ensemble avoids any artificial effect due to the thermostat and the barostat
action, important when studying dynamical properties.



\section{Results and Discussion}
\subsection{The ice polymorphs at different conditions}

To detect structural changes of the ice phases at different
thermodynamical conditions, the following three sets of (T, P) points in the phase
diagram of water were selected:
a) a set of points (16 points) located near the middle of the stability region of
   each polymorph (as explained later, for ice IV and VI it was necessary to modify
   the initial guess of the stable zone);
b) a point at T=$150$ K and P=$1$ MPa;
c) a point at T=$150$ K and P=$1$ TPa. 
For the b) and c) points all sixteen ice structures are simulated.
In the following, the a) set will be usually referred to as ``native
conditions'', the b) point as ``low pressure'' and the c) point as ``high pressure''.

In Table \ref{ices} the last three columns contain the energy difference between 
every structure and the most stable one, which happens to be always ice XI for
the three force fields considered\cite{vega-gdr1, vega-gdr2}. Absolute values
would have been difficult to directly compare because RWFF contains the 
intramolecular bond energy which is naturally not present for TIP4P and SPC/E.
The relative energy ordering for RWFF is not so different from the other two
force fields and we immediately note that, apart from ice XI (the lowest
configurational energy structure), the second lowest energy is cubic ice and
the others follow a similar trend. The energies for high pressure ices (V, VI,
VII, VIII) show very high values for RWFF with respect to the rigid models.
This is explained by the fact that O-H intramolecular distances are shorter
and this part of the interaction dominates in compressed water molecules.
This is confirmed by an analysis of the radial distribution functions.
Ice X shows an extremely high $U_{pot}$ value because the H atoms stay in a
shared lattice position between two oxygens.


\subsection{Native conditions}
\label{equilibration}
Following the MD procedure described above, the relaxation of every ice structure
has been performed within the characteristic region of stability on the phase
diagram (Figure \ref{phase-diagram}). The corresponding values of temperature
and pressure, see Table \ref{conditions}, have been taken approximately in the
middle of each stability zone.
With this choice, the powder diagrams of the equilibrated structures have then been
calculated, and  each powder diagram has been compared with the one
of the corresponding ideal structure reported in Table \ref{ices}, in order to reveal 
possible structural changes.
The powder diagrams for most polymorphs were found substantially identical for both
the initial configuration and the final configuration after equilibration.

\begin{table}
\center
\caption{Temperature and pressure of the chosen thermodynamics states}
\begin{tabular}  {|l|l|l|l|l|l|l|l|l|l|l|l|l|l|l|l|l|}			
\hline
structure  & Ic & Ih & II& III& IV& V& VI& VII\\
\hline
T (K)       & 130& 250& 200& 250& 110& 250& 225& 350 \\
\hline
p (GPa)    & $10^{-7}$&$10^{-7}$ & 0.30& 0.30& $10^{-7}$ & 0.53& 1.50& 10.00  \\
\hline
\hline
structure  & VIII& IX& X& XI& XII& XIII & XIV& XV \\
\hline
T (K)       & 150& 50& 150& 50& 180& 130& 120& 100 \\
\hline
p (GPa)    & 10.00& 0.30& 100.00& $10^{-7}$& 0.81& 0.50& 1.20& 1.10 \\
\hline
\end{tabular}
\label{conditions}
\end{table}

\begin{figure}
\centering
\includegraphics[width=1\textwidth]{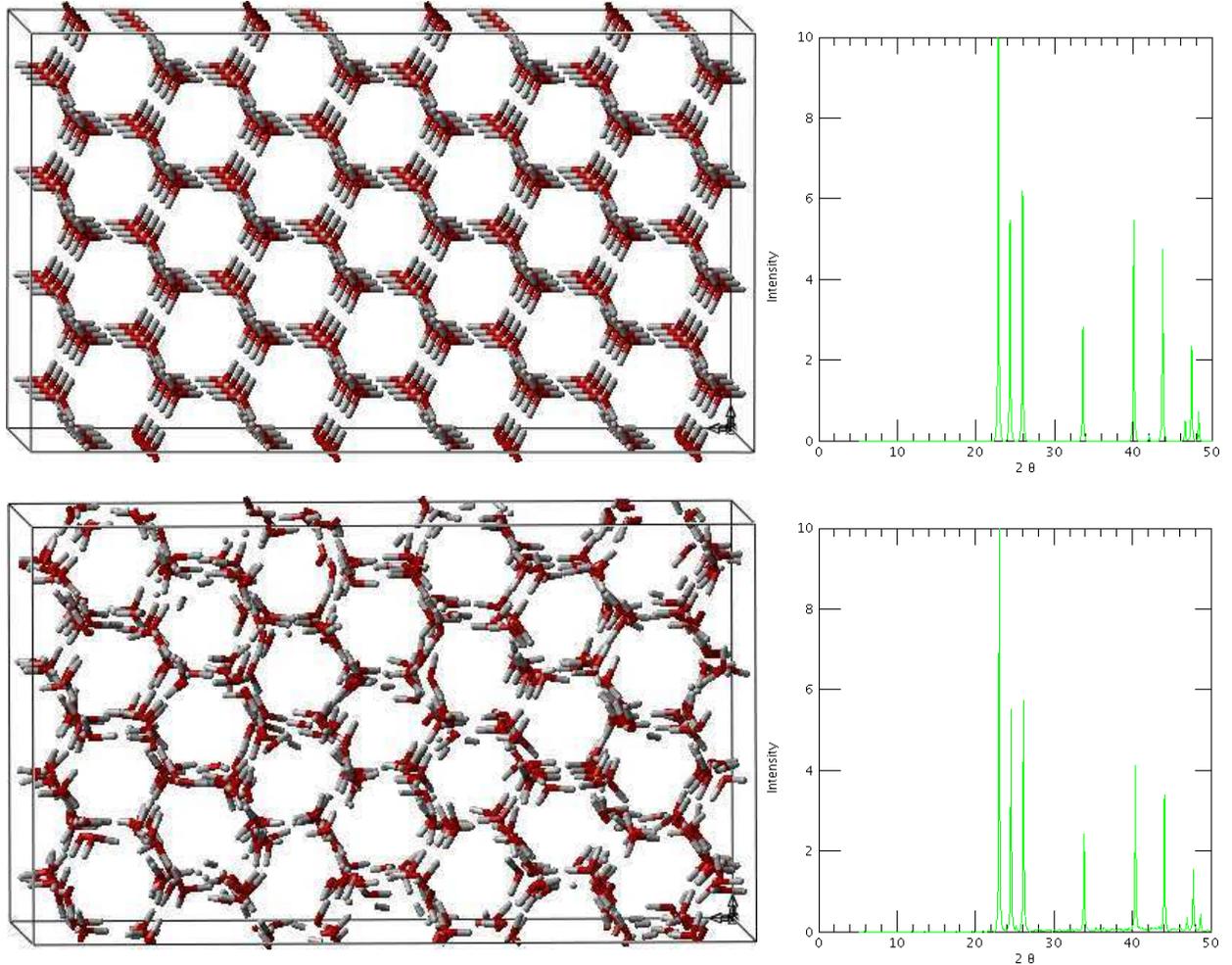}
 \caption{Structure snapshot and powder diagram of ice Ih for the ideal
 structure (top) and after thermalization at $T=250$ K and $P=1$ MPa (bottom)}
\label{Ih}
\end{figure}

The only difference observable between the initial and post MD powder diagrams
is the background noise. In fact, the equilibrated structures differ slightly 
from the ideal ones as a result of small lattice deformations at different length
scales. We observed two different effects of this phenomenon: most structures
show just thermal noise with average positions equal to the ideal ones, on the
other hand, a few of them (i.e. ice V and XIII in Fig. \ref{powder_lowPressure}) 
thermalize into imperfect lattices. This (stable) distortion is easily detectable
in the powder diffraction diagrams as a low but broad bulge centered on the main peaks.
On the other hand, as an example of pure thermal lattice distortion, Fig. \ref{Ih} shows
that for ice Ih the noise has a very low intensity, it is more uniformly distributed
over the powder diagram and should average out to zero in the limit of an infinite system.


The difference between the initial and the post MD structures is quantified by
the distance in average linkage, graphically represented in the dendrograms. This
adimensional quantity should be used as a relative similarity index: the nearer to
the left on the dendrogram two structures are connected, the higher will be their
structural similarity.

\begin{figure}
\centering
\includegraphics[width=0.5\textwidth, angle=270]{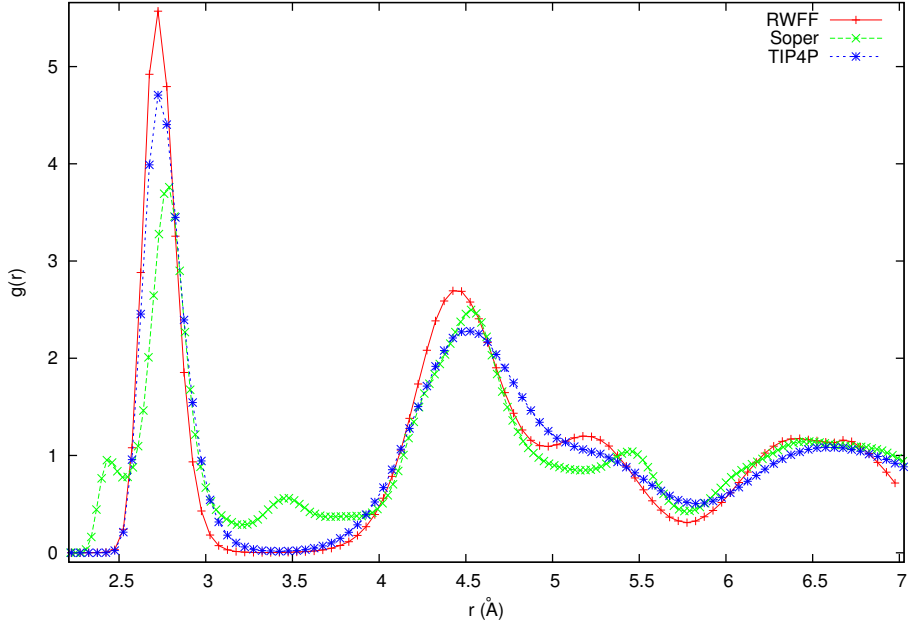}
 \caption{Oxygen radial distribution function for ice Ih after thermalization at
 $T=250$ K and $P=0.1$ MPa. Experimental data ($T=220$ K) from Soper (green) is
 compared to TIP4P (blue) and to the RWFF (red) used in this work.}
\label{Ih-gdr}
\end{figure}

As an example, the pictures of ice Ih before and after equilibration are shown 
in Fig. \ref{Ih}
(left part: top and bottom).
At $0$ K the structure is ideal and the powder diagram is free of noise (top
right of Fig. \ref{Ih}). The snapshot of the structure at $250$ K shows some random
molecular displacements. As can be seen in the direct space as well as in the
reciprocal space (bottom right), the structure remains stable.
To further check the geometrical structure of ice Ih after the MD at $T=250$ K 
and $P=0.1$ MPa, we plotted in Fig. \ref{Ih-gdr} the oxygen-oxygen radial
distribution function for the RWFF, for the experimental data (Soper\cite{soper})
and for TIP4P (taken from Vega et al.\cite{vega-gdr1}.) It should be noted that
Soper data was obtained at $T=220$ K.
The most striking difference between the MD curves and the experimental one is
the presence of two small peaks, one on the left and one on the right, of the
first big peak at about $2.8$ \AA. This value for the nearest neighbor distance
agrees well with the experimental one reported by Kuhs and Lehmann\cite{kuhs-lehmann}.
Moreover, the second peak of the radial distribution function, from which we can
compute the O-O-O angle, corresponds to an angle of about $107$ degrees, which is
also in good agreement with the data reported in the previously cited reference
\cite{kuhs-lehmann}.
A detailed comparison such as the one reported in Morse and Rice \cite{morse-rice}
would be interesting, but it is beyond the scope of this paper.
The agreement between our data and the experimental one is of the same degree of
accuracy as TIP4P or even better. Since the RWFF was obtained by fitting
exclusively experimental liquid water radial distribution functions, we believe
that the model is capturing a fair amount of water physics with a spherically symmetric
interaction. We should recall once more that RWFF is totally flexible having no
proper ``bonded interactions'' versus the constrained geometry of the other models.
This fact makes even more remarkable the fact that we are able to simulate stable
complex structures such as ice X for such long times.

From this systematic analysis it turns out that the reactive water force field
(RWFF) generates stable ice structures under experimental conditions with the 
exception of ice IV and VI. Ice IV and VI collapsed into amorphous states due to
the fact that their stability regions are very narrow and not well defined, making
it difficult to choose the reference thermodynamic points characterizing such phases.
In the paper by Vega et al.\cite{vega-gdr1} the used temperatures and pressures
to simulate these two structures are different from experimental values and it
turned out that, by using their pressure values (for ice VI we had to increase 
pressure by $0.4$ GPa with respect to the experimental value, while for ice IV
pressure was set to $0.1$ MPa), stable structures were achieved with both
isotropic and anisotropic barostats.
Please note that T and P reported in Table \ref{conditions} are those for which
we obtained stable structures.

\subsection{Low pressure}

The second batch of MD simulations has been performed slightly above atmospheric
pressure at $P=1$ MPa and $T=150$ K, always following the MD procedure of Section
\ref{protocol}. In such a thermodynamic point, all sixteen crystal structures
have been studied.

\begin{figure}
\centering
\includegraphics[width=0.6\textwidth]{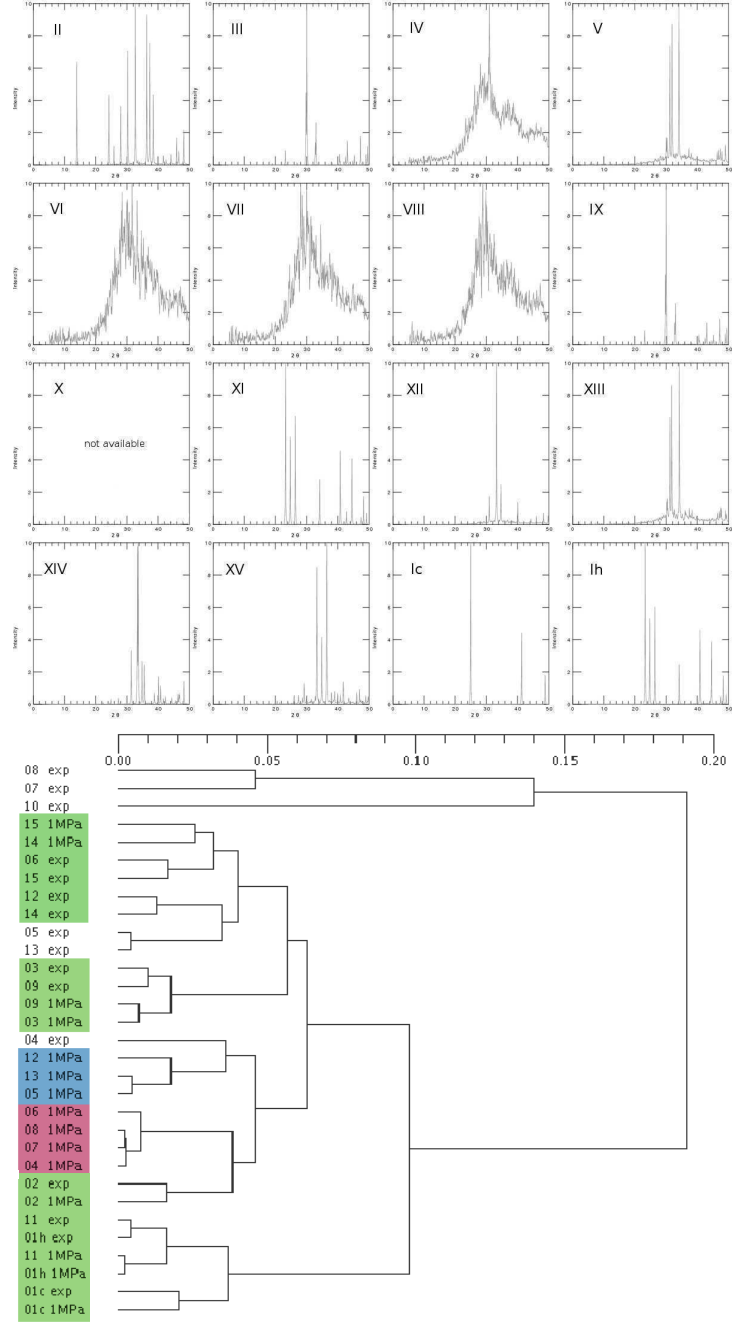}
\caption{Simulated powder diagrams of the final crystal structures at $T=150$ K
and $P=1$ MPa and their related dendrogram. Green indicates stability,
blue stands for stability with some crystal distortion and red for total
amorphization of the structure. MD simulation of ice X at this pressure was not
possible. Ideal structures coming from diffraction experiments are also included
to show the pairings with simulated crystals.} 
\label{powder_lowPressure}
\end{figure}

Figure \ref{powder_lowPressure} shows the powder diagrams of the
final configurations (upper part) along with the ideal structures inferred
from neutron diffraction experiments (labeled as ``exp'').
The low pressure ices (Ic, Ih, XI) remain stable for the full simulation time of
$200$ ps. The powder diagram of all high pressure ices (VII, VIII, X) loses any
significant peak. The structures are unstable under low pressure and transform
into amorphous ice, well reproducing the experimental results.

The ``medium pressure'' ices react in various ways, but
to a different degree depending on their position on the phase diagram.
The structures of II, III, and IX, not too far from their stability
regions in the phase diagram, were stable during the entire simulations.
The ices V, XII, XIII develop some distortions in the crystal structure, which
are reflected by an increased background noise in their powder diagrams.
The modification VI, whose stability region occurs at a higher pressure, and
the metastable modification IV, collapse completely.

The dendrogram shown in Fig. \ref{powder_lowPressure} (lower part) has been
generated by applying the clustering method to the whole set of initial and final
configurations, and offers a quick summary of the whole set of behaviors described
above.

The cluster can be divided in three different categories: `amorphous',
`stable with some deformation' and `stable' crystal structures (respectively red, 
blue and green in Fig. \ref{powder_lowPressure}). 

All amorphized structures (IV, VI, VII and VIII) form one single
subcluster with a tiny distance measure ($\sim 0.01$), indicating that all of the
structures are more or less identical. 

All partially deformed crystal structures (V, XII and XIII) belong to a different
subcluster. However, their relative distances within the dendrogram are larger than
those seen within the amorphized (red) subcluster, indicating a higher differentiation
among the structures. Distance between ice V and XIII is small due to the known
ordered/disordered proton structure relationship. Indeed, all three powder diagrams
maintain the significant peaks inherited from the original structures. 

For all the stable crystal structures,
the final and the initial configurations turn out to be very similar, but in the
case of ice crystal couples which only differ for the proton order/disorder status 
(Ih/XI and III/IX), initial and final samples cannot be distinguished and belong
to a single subcluster.

Simulation of ice X was not possible at low pressure due to the huge forces at the
beginning of the simulation not being compensated by external pressure.

\subsection{High pressure}

\begin{figure}
\centering
\includegraphics[width=0.8\textwidth]{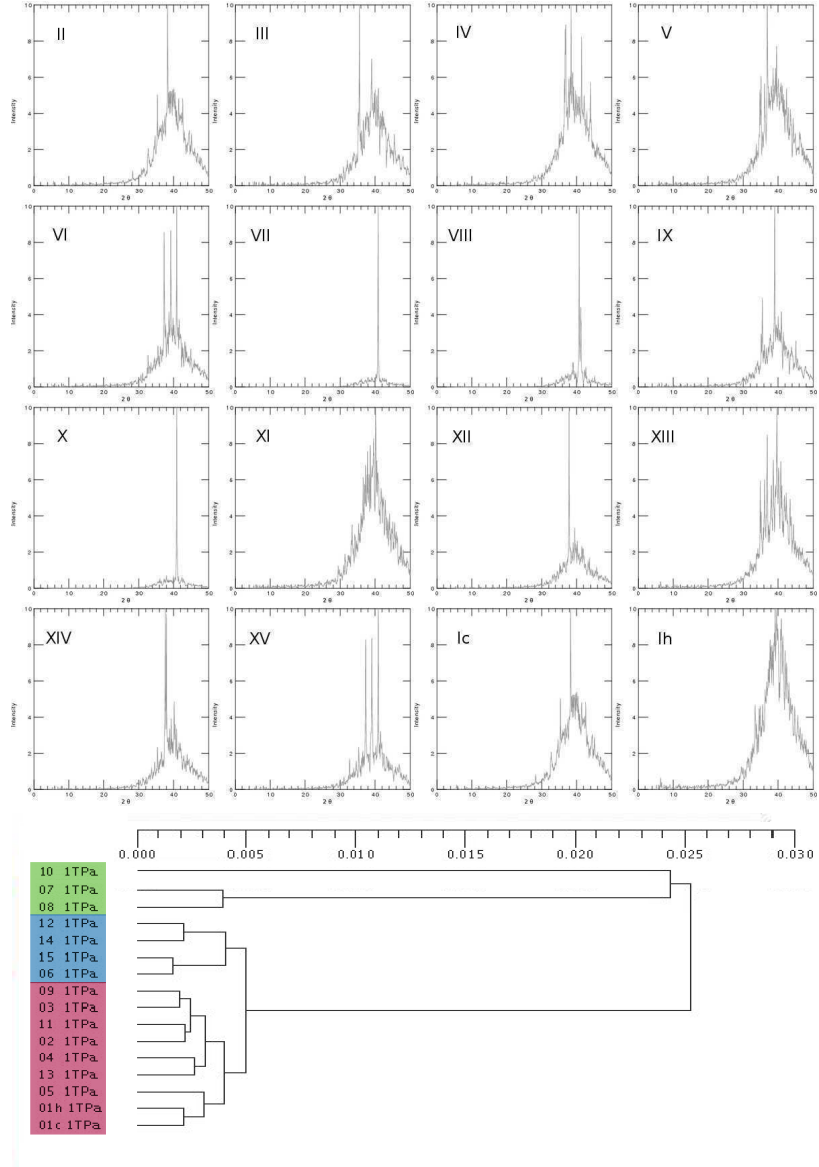}
\caption{Simulated powder diagrams of the final crystal structures at T=$150$ K
and P=$1$ TPa and the related dendrogram. Green indicates stability,
blue stands for partial stability and violet total amorphization of the structure.}
\label{powder_highPressure}
\end{figure}

The third batch of MD simulations was calculated at $T=150$ K and $P=1$ TPa.
Also in this thermodynamic point, all sixteen crystal structures have been
studied. The results are summarized in Figure \ref{powder_highPressure}.
It should be noted that the scale used in this dendrogram is much smaller with
respect to the low pressure dendrogram because more structures collapse (so
becoming all quite similar) and also because we decided not to respresent the
perfect theoretical structures.
The high pressure polymorphs of ice, VII, VIII, and X, turn out to be
stable even if some slight deformation of the crystals is observed. All of the
relevant peaks are retained nonetheless. The other structures fall into two
clusters. The first one contains
partially amorphized crystals (ices VI, XII, XIV and XV), while the other one
contains structures which became almost completely amorphous (ices Ic, Ih and XI).
From Fig. \ref{phase-diagram}, it is clear that all polymorphs belonging to
the second cluster are stable at low pressure while, and as already observed for
the low pressure simulations, the medium pressure polymorphs behave in a mixed
way. The remaining ice structures, II, III, IV, V, IX and XIII, collapse
completely during the simulation.

\section{Conclusions} 
The present study was able to show that, within a proper force field parameterization,
classical Molecular Dynamics can be successfully used to simulate the
temperature and pressure dependence of static properties of ice
crystals, such as structure, proton order-disorder relations and phase
stability.

The success is linked to the use of the RWFF potential which, at the moment,
appears to be one of the most powerful potentials for the description 
of water properties over an extended range of thermodynamic conditions,
as well as in different complex systems such as, for example, hydrated acid polymers.

The success of the approach is shown through the use of a computational
crystallography protocol, which is based on MD simulations, powder diffraction
diagram determination and cluster analysis.

The proposed protocol appears to be powerful, stable and well suited for
automated crystal structures determination.



%
\end{document}